# Graph-Coupled HMMs for Modeling the Spread of Infection

**Wen Dong, Alex "Sandy" Pentland**
M. I. T. Media Laboratory
M. I. T.
{wdong, sandy}@media.mit.edu

**Katherine A. Heller**
Department of Brain and Cognitive Sciences
M. I. T.
{kheller@gmail.com}

## Abstract

We develop Graph-Coupled Hidden Markov Models (GCHMMs) for modeling the spread of infectious disease locally within a social network. Unlike most previous research in epidemiology, which typically models the spread of infection at the level of entire populations, we successfully leverage mobile phone data collected from 84 people over an extended period of time to model the spread of infection on an individual level. Our model, the GCHMM, is an extension of widely-used Coupled Hidden Markov Models (CHMMs), which allow dependencies between state transitions across multiple Hidden Markov Models (HMMs), to situations in which those dependencies are captured through the structure of a graph, or to social networks that may change over time. The benefit of making infection predictions on an individual level is enormous, as it allows people to receive more personalized and relevant health advice.

## 1 INTRODUCTION

Growing amounts of information available from social network data afford us an unprecedented opportunity to answer questions about the individual agents who are represented in these social networks, typically as nodes, via the use of statistical models. The goal of this paper is to provide a framework for modeling the dynamical interactions between individual agents in a social network, and to specifically apply it to modeling the spread of infection.

We present the Graph-Coupled Hidden Markov Model (GCHMM), a discrete-time model for analyzing the interactions between agents in a dynamic social network.

New data and computational power have driven recent developments in models of epidemic dynamics, in which individuals in a population can express different epidemic states, and change state according to certain events. These dynamics models range from assuming homogeneous individuals and relations [Kermack and McKendrick, 1927] to incorporating increasingly more information on individuals and their relationships [Eubank et al., 2004, Salathé et al., 2010, Stehlé et al., 2011, Hufnagel et al., 2004]. For example, Eubank [Eubank et al., 2004] predicted outbreaks ahead of time by placing sensors in key locations, and contained epidemics through a strategy of targeted vaccination using land-use and population-mobility from census data. Salath [Salathé et al., 2010] looked at matching absentee records at a high school with the epidemic size predicted by the susceptible-exposed-infectious-recovered (SEIR) model.

In this paper, we focus on modeling the spread of infection at an individual level, instead of for an entire population. We aim to predict the state of each individual's health, and the paths of illness transmission through individuals in a social network. This information helps us to understand how disease is transmitted locally within a network, and to give individual-level behavioral advice about how to best maintain good health.

The social network data that we employ for this research is mobile phone data collected from 84 people over an extended period of time. The data track each person's proximity to others within the network, and daily symptom reports signal whether individual members of the network might be ill.

Our graph coupled hidden Markov model (GCHMM) incorporates dynamic social network structure into a coupled hidden Markov model [Brand et al., 1997]. A GCHMM allows us to predict at an individual level how the spread of illness occurs and can be avoided. Our results point to a new paradigm of infection control based on sensor networks and individual-level modeling.

Access to dynamic social networks is an essential part of modeling the spread of disease, and is useful for many other real-world social applications as well. The study of dynamic social networks has attracted considerable research in the machine-learning community [Goldenberg et al., 2010]. Since diffusion within dynamic social networks is important in many applications, we believe the GCHMM will also be useful for studying the dynamics of fads, rumors, emotions, opinions, culture, jargon, and so on [Castellano et al., 2009]. Even though we focus on epidemiology in this paper, the same model could be applied to determining, for example, to what extent an individual will change his opinion to match the value of a random neighbor in the formation of community consensus [Holley and Liggett, 1975], to what extent an individual will change one of his traits to match the value of a random neighbor using the Axelrod model of culture formation [Axelrod, 41], how a real-world vocabulary is formed at the society level through imitation and alignment at the individual level [Steels, 1995], and for any of these what further implications might be at both the individual level and the network level.

This paper therefore makes several novel contributions to the field of human behavior modeling and machine learning: 1) We introduce a new class of models, GCHMMs, which combine coupled HMMs with dynamic social networks. We inject dynamic social network structure into the CHMM allowing us to use this class of models for a potentially very wide range of multi-agent network or behavior modeling tasks (e.g. rumor, innovation, or organizational flow in social networks). 2) We specify a particular model in that class, which is a novel model for epidemics. This model allows us to make individual-level epidemics predictions not enabled by previous epidemics methods. 3) We provide methods for performing inference, generally in the case of 1) and specifically in the case of 2), and discuss how they relate to each other and previous work. These methods provide tools for researchers interested in a broad range of multi-agent and social network applications. 4) We validate our epidemics model on an interesting new dataset which tracks the spread of illness on a local, individual level.

The rest of the paper is organized as follows. In section 2 we review the coupled hidden Markov model. In section 3 we introduce the GCHMM for multi-agent modeling in dynamic social networks. In section 4 we show how the GCHMM can be applied to modeling the spread of infection in networks, and in 5 we derive the Gibbs sampler for parameter learning and latent state estimation of the GCHMM for epidemics. In section 6 we apply the GCHMM to our epidemic data from an undergraduate university residence hall, which includes daily symptom reports and hourly proximity tracking.

## 2 COUPLED HIDDEN MARKOV MODELS

A coupled hidden Markov model (CHMM) describes the dynamics of a discrete-time Markov process that links together a number of distinct standard hidden Markov models (HMMs). In a standard HMM, the value of the latent state at time $t$ ($X_t$) is dependent on only the value of the latent state at time $t-1$ ($X_{t-1}$). In contrast, the latent state of HMM $i$ at time $t$ in the CHMM ($X_{i,t}$) is dependent on the latent states of all HMMs in the CHMM at time $t-1$ ($X_{\cdot,t-1}$).

The CHMM generative model is defined as follows:

$$X_{i,t} \sim \text{Categorical}(\phi_{i,X_{\cdot,t-1}}) \quad (1)$$
$$Y_{i,t} \sim F(\theta_{X_{i,t}}) \quad (2)$$

where $X_{i,t}$ is the hidden state of HMM $i$ at time $t$, $Y_{i,t}$ is the emission of HMM $i$ at time $t$, $X_{\cdot,t-1}$ is a vector of the state values of all HMMs at time $t-1$, and $\phi_{i,X_{\cdot,t-1}}$ is a vector the dimensionality of which is equal to the number of states in the HMM and the entries of which sum to 1. The entries in $\phi_{i,X_{\cdot,t-1}}$ represent the probability that the state variable in HMM $i$ will transition from its state at time $t-1$ to each possible state at time $t$, given the states of all other HMMs at time $t-1$. $\theta_{X_{i,t}}$ is the emission parameter for observations that derive from state $X_{i,t}$. The graphical model for the CHMM can be seen in figure 1.

Historically, inference in CHMMs has been achieved via maximum likelihood estimation, usually using an Expectation-Maximization (EM) algorithm. Specialized CHMMs are often used in practice, however, because a CHMM with $M_i$ states for HMM $i$ has $\prod_i M_i$ states in total, and the state transition kernel $\phi_i$ is a $\prod_i M_i \times \prod_i M_i$ matrix, both of which are a considerable size. Many specializations either omit the inter-chain probability dependence, such as in the factorial hidden Markov model [Ghahramani and Jordan, 1997, Brand et al., 1997], or introduce fixed sparse inter-chain probability dependence, such as in hidden Markov decision trees [Jordan et al., 1996]. The dynamic influence model [Pan et al., 2012] allows one chain to probabilistically depend on all other chains through only a few sufficient statistics without increasing modeling complexity.

However, in the following sections we introduce the GCHMM, which differs from the models described above in considering HMMs that are coupled based on social network structure, and for which efficient inference can be performed based on the inter-chain "relations" of sufficient statistics.

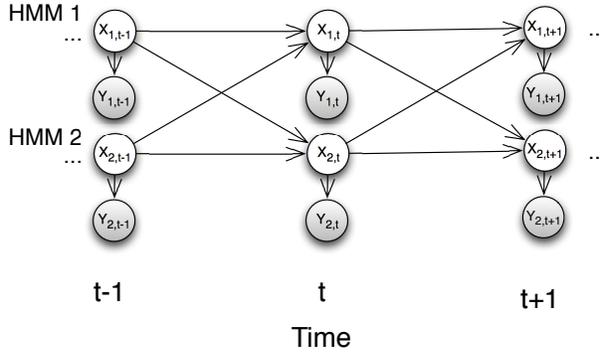

Figure 1: CHMM graphical model

The CHMM generative model can easily be formulated in a Bayesian manner, which helps to avoid overfitting problems in real-world applications [Beal, 2003]:

$$\theta_{X_i} \sim \text{Conj}(\gamma) \quad (3)$$
$$\phi_{i,X_\cdot} \sim \text{Dirichlet}(\alpha) \quad (4)$$

where Conj refers to the distribution conjugate to $F$, above, and $\gamma$ and $\alpha$ are shared hyperparameters. Here, $\theta_{X_i}$ and $\phi_{i,X_\cdot}$ are drawn once for all times $t$. $X_i$ is the set of all state values that HMM $i$ can take on, while $X_\cdot$ is the set of all state values that all HMMs can take on in the CHMM.

This is the Bayesian formulation of the CHMM that we extrapolate in the next section in order to deal with situations wherein the relationship between HMMs is mediated by a dynamic social network.

## 3 GRAPH-COUPLED HIDDEN MARKOV MODELS

Here, we introduce the graph-coupled hidden Markov model (GCHMM) to model how infection and influence spread via the interactions of agents in a dynamic social network.

Let $G_t = (N, E_t)$ be a dynamic network, where each $n \in N$ is a node in $G_t$ representing an individual agent, and $E_t = \{(n_i, n_j)\}$ is a set of edges in $G_t$ representing interactions between agents $n_i$ and $n_j$ at time $t$. Graph $G_t$ changes over time, where there is a discrete number of observations of the structure of the graph at times $t \in \{1 \ldots T\}$. Changes in $G_t$ over time represent changes in interactions between agents in the network over time.

We can use the CHMM in conjunction with dynamic network $G_t$ if we let $X_{n,t}$ be the state of agent (node) $n$ at time $t$, and $Y_{n,t}$ be noisy observations about agent $n$ at time $t$. $G_t$ restricts the connections between latent state variables in the CHMM, allowing us to model multi-agent phenomena of interest while potentially allowing for efficient inference.

The generative model of the GCHMM in its most general form is as follows:

$$X_{n,t} \sim \text{Categorical}(\phi_{n,X_{e:\{n,\cdot\}\in G_t},t-1}) \quad (5)$$
$$Y_{n,t} \sim F(\theta_{X_{n,t}}) \quad (6)$$
$$\theta_{X_n} \sim \text{Conj}(\gamma) \quad (7)$$
$$\phi_{n,X_{e:\{n,\cdot\}\in G_t}} \sim H(X_{e:\{n,\cdot\}\in G_t}, \mu) \quad (8)$$

Unlike the CHMM, whose transition matrix $\phi_n$ for HMM $n$ is dependent on all other HMMs, $\phi_n$ in the GCHMM is dependent on only the HMMs that have edges in the graph connected to node $n$. Thus, $\phi_{i,X_\cdot}$ becomes $\phi_{n,X_{e:\{n,\cdot\}\in G_t}}$. We assume that $n$ itself is also included in this set of HMMs on which $n$ is dependent. There is also a difference in the prior distribution of $\phi_n$. In the CHMM, the prior is the same for all rows of the transition matrix, whereas in the GCHMM the prior on $\phi_n$, $H$, depends on the values of the states of the HMMs on which $n$ is dependent at the previous time step.

The graphical model for the GCHMM is given in figure 2. Here, we show a GCHMM with 3 HMMs. Network structure $G_t$ is depicted in the bubbles above each time step, also showing the dependency structure corresponding to $G_t$ in the GCHMM graphical model. The structure for $G_{t-1}$ is not displayed, since it would be off the left side of the figure.

This is a discrete-time multi-agent model, and thus it approximates its continuous-time multi-agent counterpart: a Markov jump process, also called a compound Poisson process. This approximation works well only when the time step size in the discrete-time model is smaller than the average rate of interaction events. In our setting, as in many settings in which these multi agent models may be used, this is not an issue.

In the following, we describe an application of the GCHMM to fit susceptible-infectious-susceptible epidemic dynamics. Here, we assume specific forms for distributions $F$ and $H$. Much like in the CHMM, efficient inference may not always be possible in the GCHMM, but we show that efficient inference can easily be done in our specific application. We expect that efficient inference in the GCHMM will be possible for many applications, since the incorporation of social networks in the GCHMM leads to sparsity in the con-nections between latent variables, since these networks are typically sparse, and the incorporation of structure resulting from network-specific prior knowledge in $H$ can also often be leveraged. For example, as in the epidemics model, $H$ can become a simple function of a few parameters and sufficient statistics of connected states.

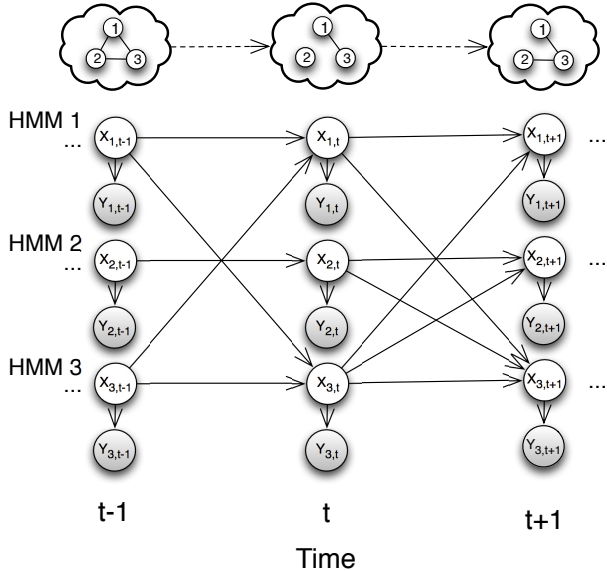

Figure 2: GCHMM graphical model

## 4 GCHMMS FOR MODELING INFECTION DYNAMICS

In this section, we show that the GCHMM can be used as a discrete-time multi-agent epidemic model, where the spread of infection is modeled on an individual level as opposed to the standard population-level models commonly used in epidemiology. In section 6, we show that the GCHMM can be applied to real-world data based on individual proximity and symptom reporting.

In particular, the GCHMM that we apply to epidemics can be seen as an individual-level version of the classic susceptible-infectious-susceptible (SIS) epidemiology model. The SIS model describes infection dynamics in which the infection does not confer long-lasting immunity, and so an individual becomes susceptible again once recovered (e.g., the common cold).

Using the GCHMM to model SIS dynamics identifies paths of infection based on individual-level interactions that are captured in our data. The classical differential equation and stochastic models of SIS dynamics work at the population level, and their variables are density and size of susceptible and infectious populations, respectively. The differential equation model $\dot{S} = -\beta \cdot SI + \gamma \cdot I$ and $\dot{I} = \beta \cdot SI - \gamma \cdot I$ for SIS specifies that the rate of change of infectious-population density is bilinear for both infectious-population density and susceptible-population density. In this system, any two individuals from the infectious population are treated as the same, as are any two individuals from the susceptible population, and therefore any two individuals from different populations have an equal chance of causing infection. The stochastic model $S + I \rightarrow 2I, rate = \beta' \cdot |S| \cdot |I|$ and $I \rightarrow S, rate = \gamma' \cdot |I|$ specifies that infection happens at a bilinear rate for both the infectious-population density and the susceptible-population density. This model enables us to reason about the randomness in the SIS system when the population size is small and randomness cannot be ignored.

The above models focus on the statistics of the spread of infection over a homogeneous population. However, we are instead interested in predicting the spread of infection on an individual level, given relevant information about each specific individual. Our goal is to explain symptom observations in a community with susceptible-infectious-susceptible dynamics at any given point in time. How likely is a person to be infectious at time $t$, given that his friends are reporting symptoms, reporting no symptoms, or not answering surveys, and given the infectious person's own survey responses and his recent proximity to his friends? Which nodes and links are critical in spreading infection in the community? How can we control infection in this community?

We use the GCHMM to address these questions, following the generative model given in section 3 and the details (state space, $H$, $F$, and so on) specified here.

$G_t = (N, E_t)$ is a dynamic network, where each node $n \in N$ represents a person in the network, and $E_t = \{(n_i, n_j)\}$ is a set of edges in $G_t$ representing the fact that person $n_i$ and person $n_j$ have interacted or come into contact at time $t$. There are two possible latent states for each person $n$ at each time $t$, $X_{n,t} \in \{0, 1\}$, where 0 represents the susceptible state and 1 the infectious state. There are six possible symptoms that can be displayed by a person at any given time, $Y_{n,t}$, which are runny nose, coughing, fever, feeling stressed, sadness, and diarrhea, and each symptom $Y_{n,t,i} \in \{0, 1\}$ represents the presence or absence of symptom $i$.

Our generative model is therefore as follows:

$$X_{n,t} \sim \text{Bernoulli}(\phi_{n, X_{e:\{n,\cdot\} \in G_t - 1}}) \quad (9)$$
$$Y_{n,t,i} \sim \text{Bernoulli}_i(\theta_{X_{n,t}}) \quad (10)$$

$$\theta_{X_n} \sim \text{Beta}(h) \qquad (11)$$

This is identical to the generative model from section 3 (the Bernoulli is the same as Categorical for 2 states), with $F$ specified as a multivariate Bernoulli distribution. Here, $h$ are given hyperparameters. The generative process for $\phi_{n,X_{e:\{n,\cdot\}\in G_t}}$ is a little more subtle, as it is defined using the interaction structure in the network, detailed below..

In keeping with the SIS model, we assume that there are certain transmission rates for the infection of one person by another, and likewise a recovery rate for an infected individual:

$$\mu = \begin{cases} \alpha \sim \text{Beta}(a,b) \\ \beta \sim \text{Beta}(a',b') \\ \gamma \sim \text{Beta}(a'',b'') \end{cases} \qquad (12)$$

$\gamma$ is the probability that a previously-infectious individual recovers and so again becomes susceptible ($p(X_{n,t+1} = 0|X_{n,t} = 1)$), $\beta$ represents the probability that an infectious person infects a previously-susceptible person ($p(X_{n_i,t+1} = 1|X_{n_i,t} = 0, X_{n_j,t} = 1, \{n_i, n_j\} \in E_{t+1})$), and $\alpha$ represents the probability that an infectious person from outside the network infects a previously-susceptible person within the network. Each of these infection probabilities is assumed to be independent. It is also assumed that a person cannot be infected by more than one infectious persons at the same time. Here, $\{a, b, a', b', a'', b''\}$ are given hyperparameters.

Therefore, we can now compute transition matrix $\phi_{n,X_{e:\{n,\cdot\}\in G_t}}$ (and thus specify $H$ from section 3) in terms of $\alpha$, $\beta$, and $\gamma$:

$$p(X_{n,t+1} = 0|X_{n,t} = 1) = \gamma \qquad (13)$$
$$P(X_{n,t+1} = 1|X_{n,t} = 0, X_{e:\{n,\cdot\}\in G_t}) \qquad (14)$$
$$= 1 - P(X_{n,t+1} = 0|X_{n,t} = 0, X_{e:\{n,\cdot\}\in G_t})$$
$$= 1 - (1-\alpha)(1-\beta)^{\sum_{e:\{n,\cdot\}\in G_t} X_{n',t}}$$

Thus our $H$ from equation 8 is now a simple function of parameters $\mu$ and sufficient statistics of connected states, $g(X)$, given by the summation term in the above equation. Intuitively, the probability of a susceptible person becoming infected is 1 minus the probability that no one, including someone from outside the network, or any of the people within the network with whom the susceptible person is known to have interacted, infected that individual. When the probability of infection is very small, it is approximately the sum of the probabilities from different sources ($P(X_{n,t+1} = 1|X_{n,t} = 0, X_{e:\{n,\cdot\}\in G_t}) \approx$ $\alpha + \beta \cdot \sum_1^{|X_{e:\{n,\cdot\}\in G_t}|} X_{e:\{n,\cdot\}\in G_t,t}$), since the probability that more than one source contributed to infection is also small.

If this assumption is correct, then the probability of seeing an entire state sequence/matrix $X$ is therefore as follows:

$$P(X, \alpha, \beta, \gamma)$$
$$= P(\alpha)P(\beta)P(\gamma) \prod_n P(X_{n,1})$$
$$\prod_{t,n} P(X_{n,t+1}|\{X_{n'\in N,t}\}, \alpha, \beta, \gamma) \qquad (15)$$
$$= P(\alpha)P(\beta)P(\gamma) \prod_n P(X_{n,1})$$
$$\prod_{t,n} \gamma^{1_{X_{n,t}=1} \cdot 1_{X_{n,t+1}=0}} \cdot (1-\gamma)^{1_{X_{n,t}=1} \cdot 1_{X_{n,t+1}=1}} \cdot$$
$$(\alpha + \beta \cdot \sum_1^{|X_{e:\{n,\cdot\}\in G_t}|} X_{e:\{n,\cdot\}\in G_t,t})^{1_{X_{n,t}=0} \cdot 1_{X_{n,t+1}=1}} \cdot$$
$$(1 - \alpha - \beta \cdot \sum_1^{|X_{e:\{n,\cdot\}\in G_t}|} X_{e:\{n,\cdot\}\in G_t,t})^{1_{X_{n,t}=0} \cdot 1_{X_{n,t+1}=0}}$$

It is assumed that all people start off in the susceptible state, and that $1_\bullet$ in the above equation is the indicator function.

In general, we can often identify $M$ types of events in agent-based models, sometimes by taking small enough time intervals in time-discretization to make first-order approximation sufficient, and count the different ways that events of type $m \in \{1, \ldots, M\}$ could change an agent's state from $x'$ at time $t$ to state $x \neq x'$ at time $t+1$, $g_m(X_{\bullet,t+1} = x, \{X_{n,t} : n, t\})$. The state transition matrix can then be defined as

$$P(X_{n,t+1} = x|X_{n,t} \neq x, X_{e:\{n,\bullet\}\in G_t}) \approx \qquad (16)$$
$$\sum_{m=1}^M \mu_m \cdot g_m(X_{n,t+1} = x, X_{n,t}, X_{e:\{n,\bullet\}\in G_t}),$$

where $\mu_m$ is the success rate of event type $m$.

This formulation of the GCHMM enables us to fit a wide range of agent-based models to "big data" sets of networked people, by specifying the right ways of counting events [Castellano et al., 2009]. For example, in the Sznajd model of opinion dynamics [Sznajd-Weron, 2005], any pair of agents with the same opinion will have a small chance $\mu$ of convincing one of their neighbors to change to their opinion. The number of ways that an individual can be convinced by a pair of neighbors is then $g(X) = \binom{\sum_{(n',n)\in G_t} 1_{X_{n'}\neq X_n}}{2}$, and the chance that an agent is convinced is $\mu \cdot g(X)$. Sznajd's model captures the intuition that we might not pay attention to a single person looking up, but we would look up if a group of people look up.

## 5 INFERENCE

In this section, we present an efficient inference algorithm for our GCHMM in order to describe and predict the spread of infection. We start by describing inference in the most general, worst case for GCHMMs, and progress from there to much more efficient inference for the epidemics model. The worst case inference algorithm for GCHMMs will be the same as for CHMMs (the case where the graph is complete). A Gibbs sampling algorithm for a particular CHMM with two chains is given in [Rezek et al., 2000], which we extend here to an unconstrained number of chains. The two sampling steps relevant to the extension to multiple chains become:

$$X_{n,t+1} \sim \text{Categorical}([\frac{p(X_{n,t+1}=j, Y|\gamma, \phi)}{\sum_k p(X_{n,t+1}=k, Y|\gamma, \phi)}])$$

$$\phi_{n,i} \sim \text{Dirichlet}([\alpha_j + \sum_{\tau=1}^{T} 1(X_{n,\tau+1} = j \land X_{\bullet,\tau} = i)])$$

Here $j$ and $k$ are states of $n$, and $i$ is the transition matrix row number corresponding to a combination of states for all nodes. We can see that this sampling procedure is not very efficient, particularly for a reasonably large number of chains.

Fortunately, in practice most social networks are sparse - far from complete. The number of parameters needed to be inferred will decrease dramatically for sparse networks (from $O(N^N)$ to $O(N^{n_{max}})$ where $n_{max}$ is the maximum number of connections for node $n$). The parameters of the transition matrix can now be sampled conditioned on the network structure:

$$\phi_{n,i} \sim \text{Dirichlet}([\alpha_j + \sum_{\tau=1}^{T} 1(X_{n,\tau+1} = j \land X_{e:\{n,\cdot\} \in G_{\tau+1}, \tau} = i)]) \quad (17)$$

Here $i$ now corresponds to a combination of the states of nodes connected to $n$ only. While significantly better than the full CHMM, this may still be intractable for large $N$ or $T$. However, the interaction structure that we are interested in is also not typically governed by unrestricted transition matrices, $\phi$. There is **structure** to interactions or behavior that we can leverage to drastically increase efficiency. One common example is that all agents may be subject to the same $\phi$. Another is that interactions themselves have a structure which can be captured by an $H(\mu, g(X))$, where

$H$ is now a function parameterized by a reasonably small set of parameters $\mu$ and sufficient statistics of the current state space $g(X)$. This allows us to sample $p(\mu|g(X))$, and then compute $\phi$, unlike in the algorithms above. This form of $H$ applies to many multi-agent network models including the emergence of collaboration, the spread of rumors and the formation of culture. It similarly applies to our epidemics model, as described in section 4, where where $g(X)$ can be efficiently computed at each node, at each time step. We now describe the resulting inference method for the epidemics model, and then briefly discuss applying the same inference scheme to some more general, $g(X)$.

The epidemics inference algorithm learns the parameters of infection dynamics, including the rate of infection and rate of recovery. It then estimates the likelihood that an individual becomes infectious from the contact with other students based on the reported symptoms of others, and even when the individual's own symptom report is not available. Finally, the algorithm enables us to make useful predictions about the spread of infections within the community in general.

We employ a Gibbs sampler to iteratively sample the infectious/susceptible latent state sequences, to sample infection and recovery events conditioned on these state sequences, and to sample model parameters.

The Gibbs sampler for the GCHMM for epidemics is given in detail below.

The Gibbs sampler takes the following as its input:

$G = (N, E)$: a dynamic network where nodes $n \in N$ represent a person in the network, and $E_t = \{(n_i, n_j)\}$ is a set of edges in $G_t$ representing that person $n_i$ and person $n_j$ have interacted or come into contact at time $t$.

$Y$: an observation matrix of symptoms indexed by time and node.

The Gibbs sampler gives the following output:

$\{X, \alpha, \beta, \gamma, \theta\}_s$: samples $s$. This includes several parameters: $\alpha$, the base rate of infection; $\beta$, the rate of infection by each infectious neighbor in $G_t$; and $\gamma$, the rate of recovery. It includes the emission matrix $\theta$, which expresses the probability of seeing each of the six symptoms given the infectious/susceptible latent state. It also includes the state matrix $X$ of the epidemics GCHMM, which shows sequences of states 0 (susceptible) and 1 (infectious) for each individual over time.

We randomly initialize the model parameters and set the state matrix so that every individual in the network is in the susceptible state. The Gibbs sampler

then iterates through sampling the latent states:

$$X_{n,t+1}|\{X,Y\}\backslash X_{n,t+1}; \alpha, \beta, \gamma \sim$$
$$\text{Bernoulli}\left(\frac{P(X_{n,t+1}=1)}{\sum_{x=0,1} P(X_{n,t+1}=x)}\right) \quad (18)$$
$$P(X_{n,t+1}=1) = P(X|\alpha,\beta,\gamma)P(Y|X)$$

where $P(X|\alpha,\beta,\gamma)$, is the same as in equation 15 minus the priors on $\alpha$, $\beta$, and $\gamma$. $P(Y|X)$ can be computed in a straightforward manner from the product of entries in the emissions matrix, $\theta$, where $P(Y|X) = \prod_{n,t,i} P(Y_{i,n,t}|X_{n,t})$.

After sampling the latent states $X$, we sample infection events. Due to the interaction structure, and to ease sampling, we introduce the auxiliary variable $R_{n,t}$ to track the source of an infection (inside or outside the network) if an infection event occurs for person $n$ at time $t+1$ (i.e., $X_{n,t} = 0$ and $X_{n,t+1} = 1$):

$$R_{n,t} \sim \text{Categorical}\left(\frac{\alpha, \beta, \ldots, \beta}{\alpha + \beta \sum_{n'} 1_{(n',n) \in E_t \cap X_{n',t}=1}}\right) \quad (19)$$

Here $R_{n,t}$ takes the value 1 if the infection event originated outside the network, and $R_{n,t} > 1$ if transmission came from someone within the network. Given the state sequences $X$ and infection events $R$, we can sample the remaining parameters from their posteriors:

$$\alpha \sim \text{Beta}(a + \sum_{n,t} 1_{\{R_{n,t}=1\}},$$
$$b + \sum_{n,t:X_{n,t}=0} 1 - \sum_{n,t} 1_{\{R_{n,t}=1\}}), \quad (20)$$
$$\beta \sim \text{Beta}(a' + \sum_{n,t} 1_{\{R_{n,t}>1\}}, \quad (21)$$
$$b' + \sum_{n,t:X_{n,t}=0;n'} 1_{(n',n) \in E_t \cap X_{n',t}=1} - \sum_{n,t} 1_{\{R_{n,t}>1\}}),$$
$$\gamma \sim \text{Beta}(a'' + \sum_{n,t:X_{n,t}=1} 1_{\{X_{n,t+1}=0\}}, \quad (22)$$
$$b'' + \sum_{n,t:X_{n,t}=1} 1 - \sum_{n,t:X_{n,t}=1} 1_{\{X_{n,t+1}=0\}}).$$

In the more general $M$ state case, we can sample $X_{n,t+1}$ from its posterior categorical distribution similar to equation 18, sample events $R_{n,t} \in \{1, \ldots, M\}$ (reflecting which type of event caused the state change of $X_{n,t}$, c.f., equation 16) similar to equation 19, and sample the success rates of different types of events similar to equation 20.

$$X_{n,t+1}|\{X,Y\}\backslash X_{n,t+1}; \mu \quad (23)$$
$$\sim \text{Categorical}\left(\frac{P(X_{n,t+1})}{\sum_x P(X_{n,t+1}=x)}\right),$$
$$R_{n,t} \sim \text{Categorical}\left(\frac{\mu_1 g_1, \ldots, \mu_M g_M}{\sum_m \mu_m g_m}\right), \quad (24)$$
$$\mu_m \sim \text{Beta}(a_m + \sum_{n,t} 1_{R_{n,t}=m}, \quad (25)$$
$$b_m + \sum_{n,t} g_m - 1_{R_{n,t}=m})$$

# 6 EXPERIMENTAL RESULTS

In this section we describe the performance of the epidemics GCHMM in predicting missing data in multiple synthetic time series, comparing to a Support Vector Machine (SVM) and standard SIS model. We also fit our epidemics model to the hourly proximity records and self-reported symptoms in a real world Social Evolution data set.

## 6.1 CONTAGION IN THE SOCIAL EVOLUTION EXPERIMENT

To demonstrate the potential of GCHMMs and our epidemics model, we use it on the data collected in the Social Evolution experiment [Dong et al., 2011], part of which tracked common cold symptoms in a student residence hall from January 2009 to April 2009. This study monitored more than 80% of the residents of the hall through their mobile phones from October 2008 to May 2009, taking periodic surveys which included health-related issues and interactions, and tracked their locations. In particular students answered flu surveys, about the following symptoms: (1) runny nose, nasal congestion, and sneezing; (2) nausea, vomiting, and diarrhea; (3) frequent stress; (4) sadness and depression; and (5) fever. Altogether, 65 residents out of 84 answered the flu surveys.

Because of the symptom reports and proximity information, the Social Evolution data is a good test bed for fitting infection models to real-world data, and for inferring how friends infect one another: For almost all symptoms, a student with a symptom had 3-10 times higher odds of having friends with the same symptom, confirming that symptoms are probabilistically dependent on their friendship network. The durations of symptoms averaged two days, and fit an exponential distribution well, agreeing with the discussed epidemic models. The base probability of a subject reporting a symptom is approximately 0.01, and each individual had a 0.006~0.035 increased chance of reporting a symptom for each additional friend with the same symptom, in line with the assumption that the rate of contagion is proportional to the likelihood of contact with an infected individual.

## 6.2 CALIBRATING PERFORMANCE

We took several steps to calibrate the performances of our epidemics GCHMM and a support vector classifier on synthetic data. First, we synthesized 200 time series – each 128 days long – from the Bluetooth proximity pattern in the Social Evolution data and different parameterizations. Then, we randomly removed the infectious/susceptible data from 10% of the population, added noise to each time series, and then inferred the held-out data with each method, for each parameterization.

The different parameterizations were (1) $\alpha = 0.01$, $\beta = 0.02$, and $\gamma = 0.3$, with observation error 0.01; (2) $\alpha = 0.01$, $\beta = 0.02$, and $\gamma = 0.3$, with observation error 0.001; and (3) $\alpha = 0.005$, $\beta = 0.045$, and $\gamma = 0.3$, with observation error 0.01. Comparing performances between (1) and (2) enables us to see the effect of observation error on algorithm performance. Comparing performances between (1) and (3) enables us to see the effect of the network on algorithm performance. Comparing performances between methods enables us to see the difference between our model-based learning and the SVM or SIS model.

We ran Gibbs samplers for 10,000 iterations, including 1000 for burn-in. We trained the SVM on a 1000-day time series synthesized using the correct parameterization, and used the number of infectious contacts yesterday, today, and tomorrow as features. We assigned different weights to the "infected" class and the "susceptible" class to balance the true and false prediction rates.

All methods can easily identify 20% of infectious cases in the missing data with little error; however, the our model-based method consistently performs better than SIS and SVM. Less noise in symptoms observations and in the contact networks of individuals significantly improves the performance of inferring missing data, as shown in Figure 3. An ROC curve shows how different algorithms compare in terms of inferring infectious cases in the held out 10% of the data.

The SVM performs poorly – especially at identifying isolated infectious cases – because it assumes that cases are i.i.d and because properly incorporating the temporal structure of epidemic dynamics into the features is not easy. The SVM also assumes that we have enough training data. This assumption often cannot be satisfied for the kinds of problems we are interested in. Lastly, we also compare to the traditional

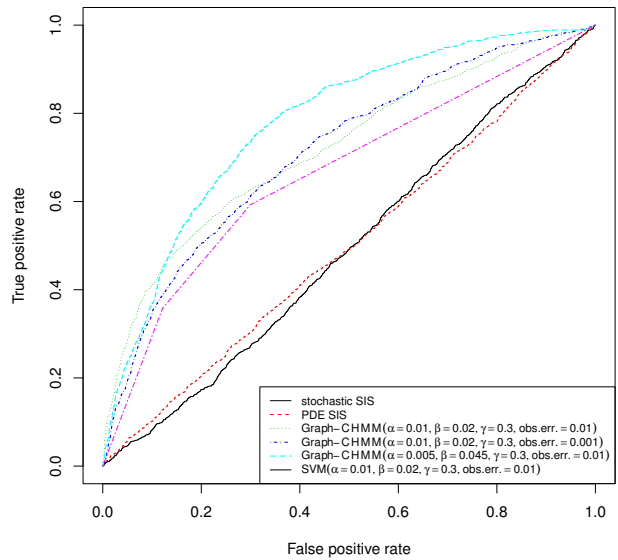

Figure 3: Less observation error (obs.err.=0.001) and better knowledge about the network ($\beta = 0.045$) result in a better trade-off between true positive rate (TPR) and false positive rate (FPR). Lack of knowledge about the network and assuming a compete graph structure result in poor trade-off between TPR and FPR. The support vector classifier has a worse trade-off between TPR and FPR than the epidemics GCHMM does.

SIS methods (both pde and stochastic). They do not predict well because they treat everyone in the population the same, regardless of individual interaction network information.

Observation noise makes inferring the individual states difficult, since it increases uncertainty in the parameters of the system. Knowledge of the dynamic contact network also affects the quality of parameter estimation and state inference. The more we know about who contacted whom, the more targeted we can be in locating the infectious cases.

## 6.3 INFERRING COLD FROM SYMPTOMS AND PROXIMITY

In this section, we report the results of our epidemics model on the Social Evolution data. In order to infer infections we extracted an hour-by-hour proximity snapshot over the 107 days that we monitored symptoms, and interpolated hourly symptom reports from the daily symptom reports. We assumed that the symptoms are probabilistically independent given the common cold (infectious) state. We ran the Gibbs sampler for 10,000 iterations, with 1000 burn-in iterations.

We do not have the clinical certainty of common cold diagnoses . However, the statistics that we discuss below give solid evidence that the epidemics model captures the structure of a infection process accompanying the symptom report.

Figure 4 shows the (marginal) likelihood of the daily common-cold states of individuals. Rows in this heat map are indexed by subjects, arranged so that friends go together, and then placed next to a dendrogram that organizes friends hierarchically into groups. Different colors on the leaves of the dendrogram represent different living sectors in the student residence hall. Columns in this heat map are indexed by date in 2009. Brightness of a heat-map entry indicates the likelihood of infectiousness - the brighter a cell, the more likely it is that the corresponding subject is infectious on the corresponding day. Sizes of black dots represent the number of reported symptoms. When a black dot doesn't exist on a table entry, the corresponding person didn't answer the survey that day.

This heat map shows clusters of common-cold infections. When larger clusters of interpersonal proximities occurred , symptom clusters lasted longer and involved more people. The heat map also tells us that subjects often submitted flu-symptom surveys daily when they were healthy, but would forget to submit surveys when in the infectious state. The Gibbs sampler nonetheless uses the data to determines that the individual was infectious for these days. Similarly, a subject sometimes reported isolated symptoms. The Gibbs sampler is able to conclude the he was most likely healthy, because the duration of the symptom reports didn't agree with the typical duration of a common cold, and because his symptom report was isolated in his contact network.

Inferred Infectious states normally last four days to two weeks. A student typically caught a cold 2-3 times during this time span. The bi-weekly searches of the keyword "flu" from January 2009 to April 2009 in Boston – as reported by Google Trends – correlated with these findings.

## 7   CONCLUSIONS

We have presented the GCHMM for modeling discrete-time multi-agent dynamics when the agents are connected through a social network. We showed that the GCHMM can be used as an individual-level SIS model to successfully predict the spread of infection throughout a social network. In the future, it would be interesting to use the GCHMM to learn graph dynamics, or to predict missing links. It would also be interesting to try to use the GCHMM in applications with a more complex transition matrix structure.

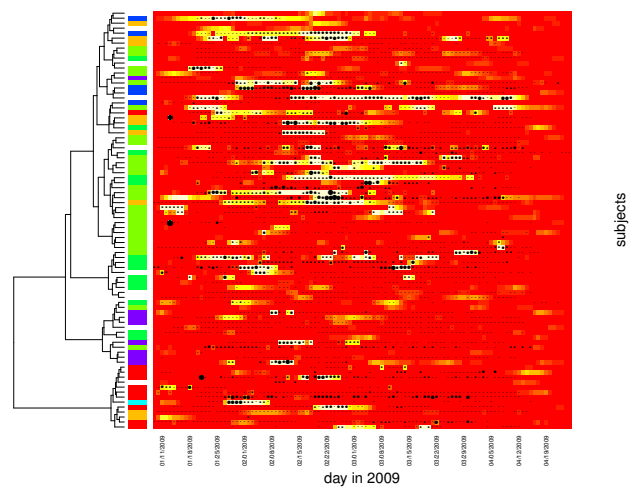

Figure 4: Our epidemics GCHMM can infer common cold state, and captures infection data from symptom self-reports and the proximity network. The size of black dots represents the number of symptoms reported, ranging from zero symptoms to all, and no black dot means no self-report.

The study of infection in small populations has important implications both for refining epidemic models and for advising individuals about their health. The spread of infection in this context is poorly understood because of the difficulty in closely tracking infection throughout a complete community. This paper showcases the spread of an infection centered on a student dormitory, with data based on daily symptom surveys over a period of four months and on proximity tracking through mobile phones. It also demonstrates that fitting a discrete-time multi-agent model of infection to real-world symptom self-reports and proximity observations gives us useful insights into infection paths and control.

**Acknowledgements**

Research was sponsored by the Army Research Laboratory under Cooperative Agreement Number W911NF-09-2-0053, and by AFOSR under Award Number FA9550-10-1-0122. Views and conclusions in this document are those of the authors and should not be interpreted as representing the official policies, either expressed or implied, of the Army Research Laboratory or the U.S. Government. The U.S. Government is authorized to reproduce and distribute reprints for Government purposes notwithstanding any copyright notation. Katherine Heller was supported on an NSF postdoctoral fellowship.

# References

Robert Axelrod. The dissemination of culture — a model with local convergence and global polarization. *Journal of Conflict Resolution*, 2(203-226), 41.

Matthew J. Beal. *Variational Algorithms for Approximate Bayesian Inference*. PhD thesis, Gatsby Computational Neuroscience Unit, University College London, 2003.

Matthew Brand, Nuria Oliver, and Alex Pentland. Coupled hidden markov models for complex action recognition. In *CVPR*, pages 994–999. IEEE Computer Society, 1997.

Claudio Castellano, Santo Fortunato, and Vittorio Loreto. Statistical physics of social dynamics. *Reviews of Modern Physics*, 81:591–646, 2009.

Wen Dong, Bruno Lepri, and Alex Pentland. Modeling the co-evolution of behaviors and social relationships using mobile phone data. In Qionghai Dai, Ramesh Jain, Xiangyang Ji, and Matthias Kranz, editors, *MUM*, pages 134–143. ACM, 2011. ISBN 978-1-4503-1096-3.

S. Eubank, H. Guclu, V. Kumar, M. Marathe, A. Srinivasan, Z. Toroczkai, and N. Wang. Modelling disease outbreaks in realistic urban social networks. *Nature*, 429: 180–4, 2004.

Zoubin Ghahramani and Michael I. Jordan. Factorial hidden markov models. *Machine Learning*, 29(2-3):245–273, 1997.

Anna Goldenberg, Alice X. Zheng, Stephen E. Fienberg, and Edoardo M. Airoldi. A survey of statistical network models. *Found. Trends Mach. Learn.*, 2:129–233, February 2010. ISSN 1935-8237.

Richard A. Holley and Thomas M. Liggett. Ergodic theorems for weakly interacting infinite systems and the voter model. *Annals of Probability*, 3(4):643–663, 1975.

L. Hufnagel, D. Brockmann, and T. Geisel. Forecast and control of epidemics in a globalized world. In *Proc Natl Acad Sci USA*, volume 101, pages 15124–9, 2004.

Michael I. Jordan, Zoubin Ghahramani, and Lawrence K. Saul. Hidden markov decision trees. In Michael Mozer, Michael I. Jordan, and Thomas Petsche, editors, *NIPS*, pages 501–507. MIT Press, 1996.

W. Kermack and A McKendrick. A contribution to the mathematical theory of epidemics. *Proceedings of the Royal Society of London. Series A, Containing Papers of a Mathematical and Physical Character*, 115(772):700–721, August 1927.

Wei Pan, Wen Dong, Manuel Cebrian, Taemie Kim, James H. Fowler, and Alex Pentland. Modeling dynamical influence in human interaction: Using data to make better inferences about influence within social systems. *Signal Processing Magazine, IEEE*, 29(2):77 –86, march 2012. ISSN 1053-5888. doi: 10.1109/MSP.2011.942737.

I. Rezek, P. Sykacek, and S. J. Roberts. Learning interaction dynamics with coupled hidden Markov models. *Science, Measurement and Technology, IEE Proceedings-*, 147(6):345–350, 2000. URL http://ieeexplore.ieee.org/xpls/abs_all.jsp?arnumber=899989.

Marcel Salathé, Maria Kazandjiev, Jung Woo Lee, Philip Levis, Marcus W. Feldman, and James H. Jones. A high-resolution human contact network for infectious disease transmission. In *Proceedings of the National Academy of Science of the United States of America*, volume 107, pages 22020–22025, 2010.

Luc Steels. A self-organizing spatial vocabulary. *Artificial Life*, 2(3):319–332, 1995. URL http://www.isrl.uiuc.edu/~amag/langev/paper/steels96aSelf.html.

Juliette Stehlé, Nicolas Voirin, Alain Barrat, Ciro Cattuto, Vittoria Colizza, Lorenzo Isella, Corinne Régis, Jean-François Pinton, Nagham Khanafer, Wouter Van den Broeck, and Philippe Vanhems. Simulation of an seir infectious disease model on the dynamic contact network of conference attendees. *BMC Medicine*, 9(1):87, 2011.

K. Sznajd-Weron. Sznajd model and its applications. *Acta Physica Polonica B*, 36(8):2537–2547, 2005.